\documentclass[twocolumn,showpacs,preprintnumbers]{revtex4}
\usepackage{graphicx}
\usepackage{dcolumn}
\usepackage{bm}
\usepackage{amsmath,amssymb}

\usepackage{multibox}

\graphicspath{{aFigures/}}

\usepackage{epstopdf}
\begin{document}

\preprint{}

\title{A periodic microfluidic bubbling oscillator: \\insight into the stability of two-phase microflows}

\author{Jan-Paul Raven}
\email{jpraven@spectro.ujf-grenoble.fr}
\author{Philippe Marmottant}
\affiliation{%
Laboratoire de Spectrom\'etrie Physique, B.P. 87,
F-38402 St Martin d'H\`eres Cedex, France
}%
\altaffiliation{CNRS - UMR 5588,
Universit\'e Grenoble I}

\date{\today}

\begin{abstract}
This letter describes a periodically oscillating microfoam flow. For constant input parameters, both the produced bubble volume and the flow rate vary over a factor two.
We explicit the link between foam topology alternance and flow rate changes, and construct a retroaction model where bubbles still present downstream determine the volume of new bubbles, in agreement with experiment. 
 This gives insight into the various parameters important to maintain monodispersity and at the same time shows a method to achieve controlled polydispersity.    
\end{abstract}

\maketitle

 
Formation and flow of bubbles in microfluidic systems attract an increasing 
attention \cite{Garstecki2004,Garstecki2005,Cubaud2004,Ganan-Calvo2001,Ganan-Calvo2004}, while potential applications cover a wide range of fields varying from biology to biomedicine (ultrasound contrast agents) to chemistry (microreactors). 
Recently these studies have been extended to the low liquid fraction case \cite{Raven2006,Garstecki2006} in which microfoams are formed. Microfoams are interesting because of the large number of interfaces that can potentially be used as a transport vector for amphiphilic molecules. Another positive aspect is the excellent control over the volume of the dispersed phase (monodispersity) with a volume variation smaller than 3 $\%$ \cite{Garstecki2004}. This is an important factor e.g. for the synthesis of new micromaterials (solid microfoams). Recently a new set of operations called `discrete microfluidics' \cite{Drenckhan2005} has showed the way to integrate microfoams in lab-on-a-chip systems. 

\begin{figure}[ht]
\setlength\unitlength{1 cm}
     \begin{picture}(7,11)
    \put(-1,4.5){\includegraphics[scale=0.45]{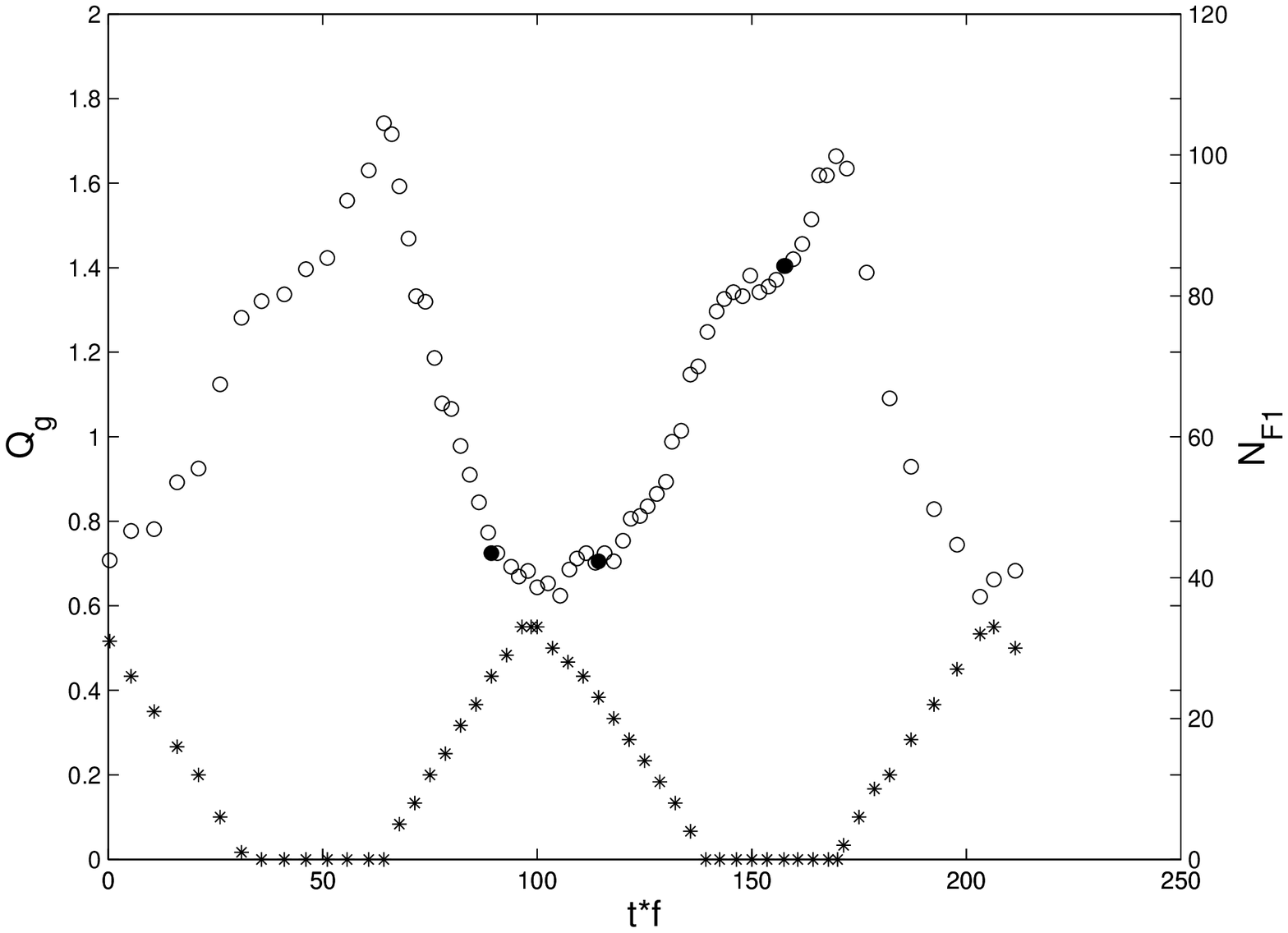}}
    \put(-1,0){\includegraphics[scale=0.35]{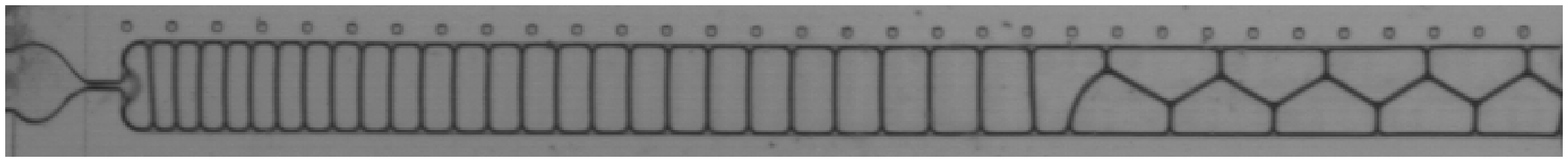}}
    \put(-1,1.5){\includegraphics[scale=0.35]{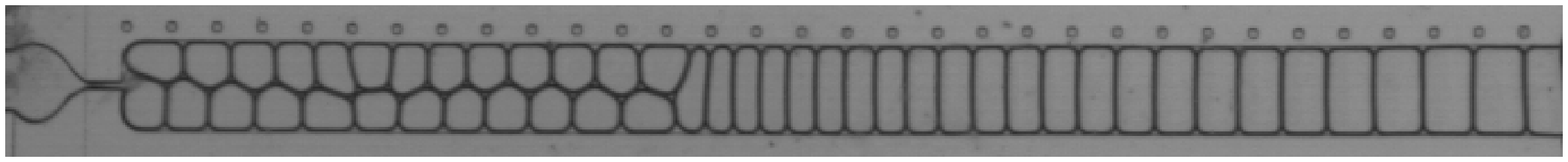}}
    \put(-1,3){\includegraphics[scale=0.35]{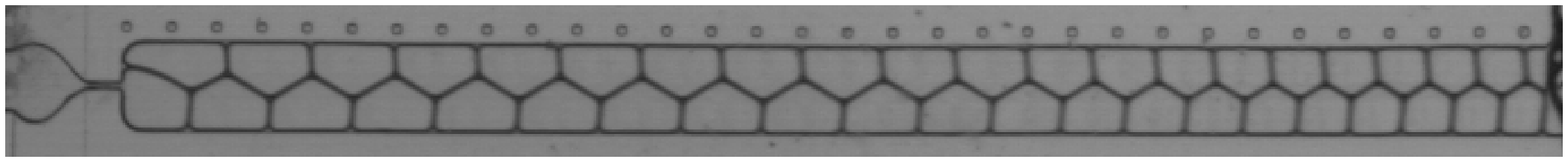}}
    \put(-1,11){(a)}
    \put(-1,1){(d)}
    \put(-1,2.5){(c)}
    \put(-1,4){(b)}
    \put(1.6,6.5){d}
    \put(2.5,7){c}
    \put(3.5,8.6){b}
    \end{picture}
	\caption{A periodically oscillating microfoam generator. a) Gas flow rate $Q_g$ (circles) ($\mu$l/s) oscillates over a factor 2.8 in time $t$ (non-dimensionalised using the bubbling frequency $f$). 
	Produced foam topology  oscillates between two states: F2 (b, whole channel, c, near entrance) and F1 (d). The number $N_{F1}$ of F1 bubbles (* in a)) is inversely correlated with $Q_g$. The liquid flow rate $Q_l=6.7\;10^{-2} \mu$l/s and gas pressure $P_g=2.90$ kPa. 
}
	\label{fig:expdata}
\end{figure}

In this letter we will show the generation of a microfoam with a periodical oscillation in its bubble volume. This is an unexpected effect in microfoam flows for which at constant input parameters both gas flow rate and bubble volume show a large periodic oscillation. We can explain it by linking bubble volume to resistance to flow of a foam in a channel. This resistance depends on the bubble volumes downstream and on the foam topology therefore the volume of a new bubble is related to the one of its predecessors. One could state that the bubble formation process has a memory. This can be used for the generation of microbubble aggregates with a controlled and potentially tunable polydispersity. It will also play an important role in defining design criteria for systems in which monodispersity is important, both for foams as for emulsions with a large viscosity contrast between continuous and dispersed phase. The theoretical model provides an effective tool to predict the  sensibility of the system to perturbations in monodispersity. It will help to select beforehand the right geometrical and experimental parameters.   


We use a conventional flow-focusing geometry \cite{Anna2003,Garstecki2004,Garstecki2005}: an inlet channel for the liquid, another one for the gas, both ending in a small orifice followed by a straight channel section (orifice width $w_{or}$ = 100 $\mu$m, channel width $w$ = 1 mm, height $h$ = 100 $\mu$m, outlet channel length  16 mm). The walls of the system are made in PDMS glued to a glass cover slide. A syringe pump (11 PicoPlus, Harvard Apparatus) is used to push the liquid (deionized water with 10\% commercial diswashing detergent Dreft, Proctor\&Gamble). For the dispersed gas phase we use nitrogen supplied from a pressurized tank via a pressure reduction valve.  The exit is at atmospheric pressure. Therefore we have access to the imposed pressure drop over the total system (orifice and outlet channel). We use a CMOS camera (F131B, Allied Vision Technologies) connected to a macro lens to capture still images and to record movies of the two-phase flow. Via image analysis, we extract the bubble volume $V_b$, bubble formation frequency $f$, from which we deduce the gas flow rate $Q_g = V_b\;f$.


Here we present a case in which for constant input parameters $P_g$ and $Q_l$ the bubble volume and gas flow rate periodically oscillate over more than a factor 2 (see Fig. \ref{fig:expdata}). A typical cycle starts with a foam consisting of two rows of bubbles (that we name F2 foam). See Fig. \ref{fig:expdata} b). The flow rate grows, and at the same time the bubble volume $V_b$ of newly formed bubbles at the orifice increases. This increase is followed by a transition of the foam topology in the channel from two rows to one row of bubbles (F2 to F1 foam) (see Fig. \ref{fig:expdata} c)). Now the foam starts slowing down while the bubble volume at the entrance is shrinking. This is followed by another transition, this time from F1 to F2 (Fig. \ref{fig:expdata} d)). When the last F1 bubbles are evacuated from the channel a new cycle can start. Over the whole cycle the bubble formation frequency $f$ stays constant. This is consistent with earlier measurements \cite{Raven2006} that showed that for bubble formation at high dispersed phase volume fraction, $f$ only depends on the liquid flow rate $Q_l$. Therefore there is a linear relation between the volume of a newly formed bubble and the velocity $v$ at which the foam flows: $V_b={Q_g}/{f}=v{S}/{f} $. Velocity is homogeneous over the channel. Image analysis shows no compressibility effects in agreement with the small pressures (typically a few kPa). The foam behaves like a plug flow. 

We stress that the bubble formation process itself is stable while the flow is not. We are in the high-pressure regime where the gas-water interface pinches off but stays in the orifice \cite{Raven2006}. 
Non-linearities due to rapid gas-water interface retraction upstream of the orifice after bubble pinch-off like in \cite{Garstecki2005b}
do not play a role. Pinch-off is stable, both in periodicity and spatial position. For a slightly lower driving pressure $P_g$, we observe a monodisperse F2-foam. A slightly higher pressure leads to a monodisperse F1-foam.

A key element for the understanding of the oscillation are the topological transitions (from F1 to F2 and \textit{vice versa}). 
The dominant force in a foam is surface tension, tending to minimize bubble surface for a given bubble volume. 
It is known that a foam is usually stuck in a local energy minimum an does not reach the global minimum \cite{Weaire1999}. The method used for the preparation of the foam is the determinant factor.
This also holds for our case in which for the same $V_b$ we can both have an F2 and an F1 foam. Furthermore the very elongated F1 bubbles are obviously far from minimising their surface to volume ratio. The formation process in this confined geometry plays an important role. 

\begin{figure}[htbp]
\setlength\unitlength{1 cm}
     \begin{picture}(7,8)
  
    \put(0,6){\includegraphics[scale=0.5]{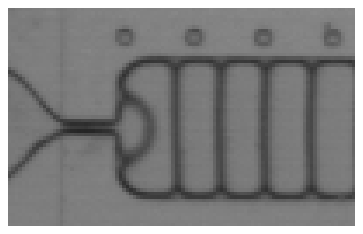}}
    \put(2,6){\includegraphics[scale=0.5]{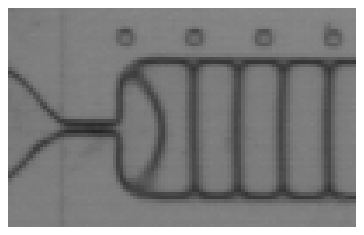}}
    \put(4,6){\includegraphics[scale=0.5]{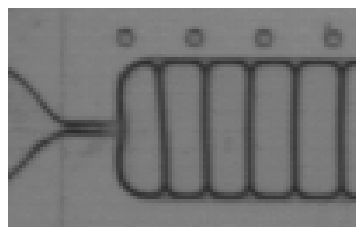}}
    \put(.2,7.25){a) F1 bubble formation}
    \put(0,4){\includegraphics[scale=0.5]{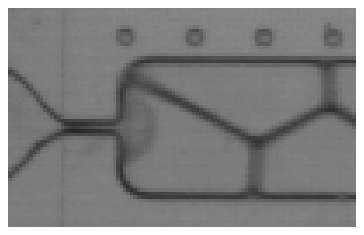}}
    \put(2,4){\includegraphics[scale=0.5]{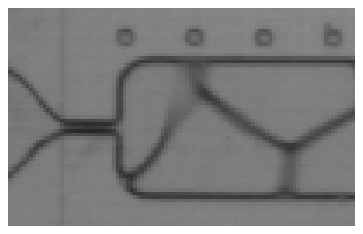}}
    \put(4,4){\includegraphics[scale=0.5]{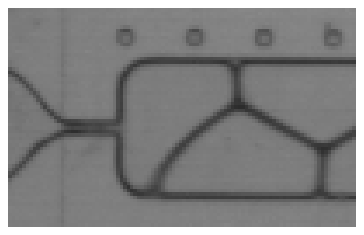}}
    \put(.2,5.25){b) F2 bubble formation}
    \put(0,2){\includegraphics[scale=0.5]{F1-180.eps}}
    \put(2,2){\includegraphics[scale=0.5]{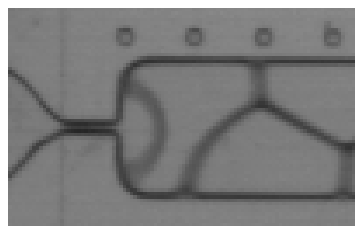}}
    \put(4,2){\includegraphics[scale=0.5]{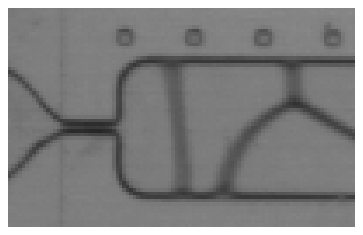}}
    \put(.2,3.25){c) Transition F2 to F1}
    \put(0,0){\includegraphics[scale=0.5]{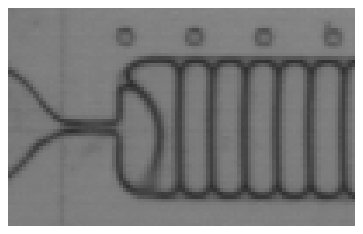}}
    \put(2,0){\includegraphics[scale=0.5]{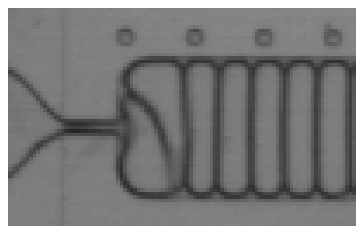}}
    \put(4,0){\includegraphics[scale=0.5]{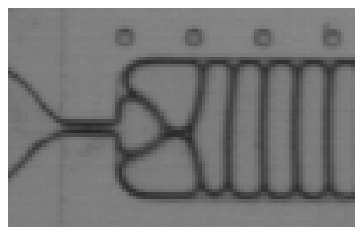}}
    \put(.2,1.25){d) Transition F1 to F2}
    \end{picture}
	\caption{Formation of the  foam topology F1 (a) and F2 (b), and the  transitions F2 to F1 (c) and F1 to F2 (d). 
}
	\label{fig:transition}
\end{figure}

We explain the topology selection mechanism by the interplay between bubble formation at the orifice and the shape assumed by the preceding bubbles. We will discuss first F1 and F2 formation, before describing transitions. In the case of the bubble formation for an F1 foam, the new bubble will first form a circular shape centered around the orifice (see pictures in Fig. \ref{fig:transition} a). A three-fold vertex (three meeting films) is formed at the point where the wall, the new bubble and the preceeding one meet. These vertices slide over the channel walls. The F2 formation is a bit more complex  (see Fig. \ref{fig:transition} b). Here again formation of a bubble starts with a spherical cap, with a vertex on each side. Because of the inherent asymmetry of an F2-foam the distance between these two vertices and the preceeding ones is unequal. Between the two vertices at shortest distance a T1-transition (bubble neighbour switching \cite{Princen1982}) 
takes place alowing the newly formed foam to relax to the F2 state. 

The transition from F1 to F2 takes place when the volume $V_b$ of the new bubble becomes that large that the inter-vertex distance is too large to allow a T1 (see Fig. \ref{fig:transition} c). We will call this volume $V_{12}$: the threshold volume for transition from F1 to F2. The opposite transition takes place at a much smaller bubble volume because of the absence of preexistent asymmetry: bubble volume has to  reduce a lot  (smaller than $V_{21}$) before the inter-vertex distance is  that small that the situation becomes unstable (probably because of interaction with previous film), that a top or bottom vertex slides faster, triggers  a T1 upon contact with the previous film, thereby completing the transition  (\ref{fig:transition} d). 

For our channel geometry the transition F1 to F2 takes place at $V_{12}/hw^{2}$ = 0.193 and the inverse at $V_{21}/hw^{2}$ = 0.536. There is a large hysteresis (a factor 2.8 between these two volumes). 
The difference in transition bubble volume probably  depends heavily on the liquid fraction as this is the determinant factor for the vertex distance at which a T1 rearrangement is triggered \cite{Princen1982,Raufaste2006}. Therefore we expect that the amplitude of the oscillation will be less pronounced for wetter foams. 

We complete the explanation of the oscillating behavior by making a link between foam topology and rheology. Driving pressure $P_g$ is related to foam velocity $v$ by the channel resistance to flow (dissipation). The foam speed is oscillating for constant driving pressure, therefore the channel resistance must oscillate as well. The principal source of dissipation for foam flow in a channel is the sliding of the liquid films between bubbles over the channel walls \cite{Cantat2004,Saugey2006}. Channel resistance scales linearly with the total length of these films in the channel. 
Orientation plays an important role. For a film of length $L$ the relevant length scale is the projected length  $L^p=L \cos \alpha$ with $\alpha$ the angle between the normal vector of the film and the flow direction. Introducing the capillary number $Ca=\mu v/\sigma$ containing the bubble velocity $v$ (estimated as $v\simeq Q_g/S$) and liquid viscosity $\mu$, the pressure drop writes
\begin{equation}
\Delta P_{channel}=\overline{\lambda}\frac{\sum L^{p}}{S}\sigma Ca^{2/3},
\label{PaCa}
\end{equation}
with $S=hw$ the channel cross section area, and $\overline{\lambda}$ a numerical constant \cite{Cantat2004, Denkov2005, Denkov2006}.

Using geometrical considerations we can calculate the film length per bubble, projected on the direction normal to flow, for the two distinct topologies: 
	\begin{eqnarray}
	L^p_{F1}&=&2w, \nonumber \\
	L^p_{F2}&=&\left(1+\frac{1}{\sqrt{3}}\frac{V_b}{hw^2}\right)w,
	\end{eqnarray}
taking into account the bottom and top wall of the channel, and neglecting the side walls. 
The largest possible bubble volume in the F2 topology is $V_b/hw^2=\sqrt{3}$. Larger volumes would violate Plateau's law saying that three films must always meet at equal angles.
Therefore for all possible bubble areas $L^p_{F1} > L^p_{F2}$. 

Summing  over the projected lengths of all the bubbles in the channel gives the total resistance prefactor  and thereby the capillary number $Ca$ (and gas flow rate $Q_g$) at a given applied pressure drop, from equation \ref{PaCa}. 

To model the foam flow we proceed in the following way. The bubbling frequency $f$ being constant whatever the bubble size, we predict the volume of a new bubble to be $V_b=Q_g/f$. Using equation \ref{PaCa}, we obtain the gas flow rate $Q_g$ as a function of the bubbles downstream in the channel,  and thus the next bubble volume,  at a given discretised time  $t f=n$, from
\begin{equation}
\overline{V_b}^{(n+1)}=\frac{1}{C\!a_f}\left(\frac{2\overline{P}}{\overline{\lambda}\;\overline{R_{total}} }\right)^{3/2},
\label{AR}
\end{equation}
with the following dimensionless quantities: $\overline{V_b}=V_b/hw^{2}$ the bubble volume, $\overline{R_{total}}=\sum_{k=n-N_b}^{n}L^{p(k)}/w$ the total resistance of the films, $C\!a_f={\mu w f }/{\sigma}$ and $\overline{P}={\Delta P_{channel} h}/{2\sigma}$ the experimental parameters  related to frequency and pressure.  The sum of the projected lengths $\overline{R_{total}}$, takes into account the $N_b$ bubbles in the channel. In addition, we enforce foam topology  transitions  at the bubble volumes $V_{12}$ and $V_{21}$, which completes the model. Essentially we calculate $\overline{V_b}^{(n+1)}$ by numerical integration over all the bubbles in the channel. Therefore a bubble influences the volume of its successors during its presence in the channel, so while travelling from the orifice to the channel exit. Thereby creating a retroactive effect.  

\begin{figure}[htbp]
\setlength\unitlength{1 cm}
     \begin{picture}(7,8.5)
    \put(-1,4){\includegraphics[scale=0.7]{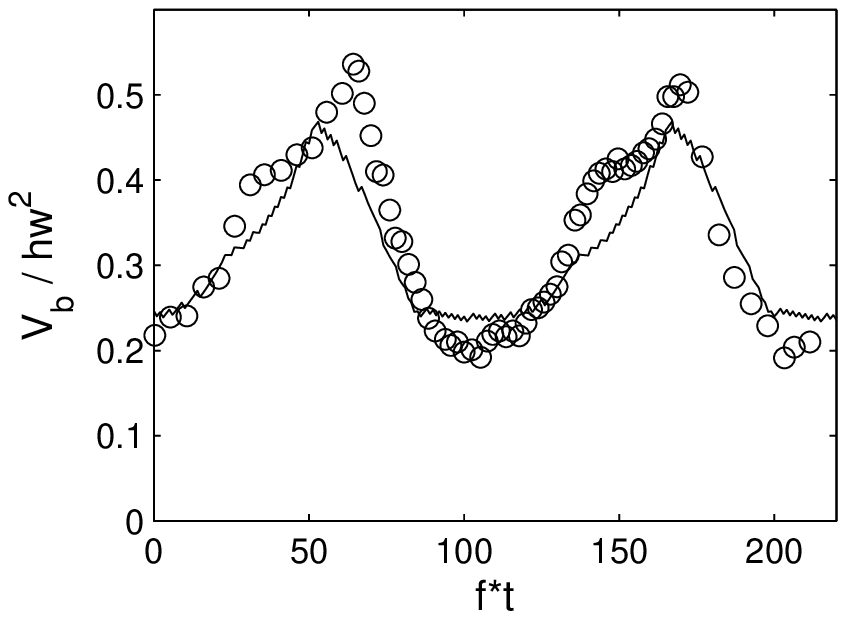}}
      \put(-1,-0.1){\includegraphics[scale=0.7]{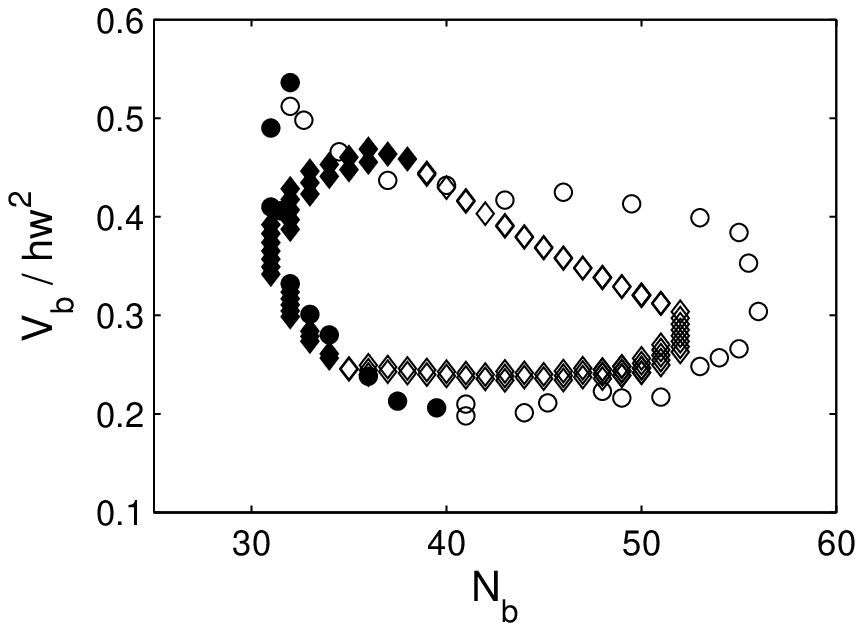}}
        \put(-1,4){(b)}
        \put(-1,8){(a)}
      
    \end{picture}
	\caption{a) Comparison between experimental data (circles) and the model (continuous line) for the bubble volume $V_b$ at the entrance of the channel. Time is normalised by the bubbling frequency $f$.
	 b) Bubble volume against the total number of bubbles in the channel $N_{b}$. ($\diamond$): model, (o): experiment. Filled symbols indicate that newly formed bubbles form an F1 topology, open symbols the generation of F2. 
Parameters $\overline{\lambda} = 67$, $\overline{C\!a_f}=8.9\;10^{-4}$, $\overline{P}=8.2$, $V_{12}/hw^{2}=0.25$, $V_{21}/hw^{2}=0.45$. 
}
	\label{fig:model}
\end{figure}

A comparison between experiment and model is shown in Fig. \ref{fig:model} confirming that we capture the essential of the physics. The model reproduces the oscillation frequency, the gas flow rate and number of bubbles in the channel nicely. 

We would like to acknowledge support and fruitful discussion with Fran\c cois Graner. We thank Gerrit Danker for constructing discussions.


\end{document}